# Helicity Maximization of Structured Light to Empower Nanoscale Chiral Matter Interaction

Mina Hanifeh[1], Mohammad Albooyeh, and Filippo Capolino[2]

*Department of Electrical Engineering and Computer Science, University of California, Irvine, California 92697, USA*

## ABSTRACT

Structured light enables the characterization of chirality of optically small nanoparticles by taking advantage of the helicity maximization concept recently introduced in[1]. By referring to fields with nonzero helicity density as chiral fields, we first investigate the properties of two chiral optical beams in obtaining helicity density localization and maximization requirements. The investigated beams include circularly polarized Gaussian beams and also an optical beam properly composed by a combination of a radially and an azimuthally polarized beam. To acquire further enhancement and localization of helicity density beyond the diffraction limit, we also study chiral fields at the vicinity of a spherical dielectric nanoantenna and demonstrate that the helicity density around such a nanoantenna is a superposition of helicity density of the illuminating field, scattered field, and an interference helicity term. Moreover, we illustrate when the nanoantenna is illuminated by a proper combination of azimuthal and radially polarized beams, the scattered nearfields satisfy the helicity maximization conditions beyond the diffraction limit. The application of the concept of helicity maximization to nanoantennas and generating optimally chiral nearfield result in helicity enhancement which is of great advantage in areas like detection of nanoscale chiral samples, microscopy, and optical manipulation of chiral nanoparticles.

## I. INTRODUCTION

A variety of natural and artificial materials are chiral, meaning their building blocks (the constituent inclusions), i.e., molecules and meta-atoms, are not superimposable to their mirror images [2]. These materials have a broad range of applications in sciences and technologies in areas like chemistry, biology, medicine, and pharmacology[3]–[5]. Although each pair of mirror-imaged molecules or other natural or engineered nanoparticles, called enantiomers, have identical elements with the same constitutions, they show different *optical* properties[6]. Conventionally, chirality is determined through circular dichroism (CD) since the *chiral* materials interact differently with circularly polarized plane waves of opposite handedness[7]. The interaction is inferred by transmitted or absorption power when the chirality of a slab or bulk of a material is being investigated, respectively [8]–[11]. However, if instead of considering optically large chiral materials we consider optically small chiral samples (with sizes smaller or comparable to the illumination wavelength), CD experiments are not anymore suitable for chirality characterization since the light-matter interaction becomes weak and the detection range of power detectors may not cover such data. Although many studies have been performed to enhance the range of chirality detection, e.g., by introducing the concept of super-chiral light which were realized by the use of counter-propagating plane waves with circular polarization of opposite handedness [12] or by the use of near-fields of plasmonic structures [13]–[23], there is not yet a platform which determines a reliable theoretical background for the use of general electromagnetic fields in detection/characterization of chirality, not to say the best field types that enhance light-chiral matter interaction. Therefore, we found it essential to provide a suitable theoretical platform for enhancing the possibility of detection and characterization of chiral nanoparticles (NPs) when using general electromagnetic fields.

In a separate work (see Ref. [1]), we have introduced the concept of helicity maximization and have elaborated that the chirality characterization of NPs via dissymmetry factor *g* [24] is achieved only when fields with maximum helicity density are employed. In that work, we have proved that such fields satisfy a universal relation between their helicity and energy densities, and have shown how this relation enables the characterization of NPs' chirality through the dissymmetry factor *g*.

In this complementary work, we study in detail the properties of two illustrative examples of structured lights, i.e., an optical beam and the nearfield of a nanoantenna, both with maximized helicity density. We demonstrate how a properly engineered nanoantenna substantially improves the chirality detection by locally maximizing the helicity and energy densities, while enabling chirality characterization. By suggesting the proposed method to employ highly localized electromagnetic fields, our findings provide a strong theoretical foundation in the study of nanoscale chirality detection and characterization of NPs using near fields and structured light.

---

[1] M.hanifeh@uci.edu
[2] F.capolico@uci.edu

## II. THEORETICAL ANALYSIS

This section is devoted to developing a theoretical basis for chirality detection and characterization of NPs by utilizing structured lights. Before starting the discussion, we emphasize to distinguish between *detection* and *characterization*: while in detection our goal is to determine whether a particle is chiral, by characterization we aim at the determination of the exact strength of the particle's chirality.

### A. Interaction of light with a chiral nanoparticle

We consider NPs whose optical response is well approximated by the fundamental electric and magnetic dipoles, and higher order multipoles are considered negligible, which is a judicious assumption for optically small NPs [25]. Moreover, we assume that the particle is isotropic, reciprocal, and is surrounded by an isotropic non-dispersive medium, and hence, the electric **p** and magnetic **m** dipole moments are linearly related to the *local* electric **E** and magnetic fields via the polarizabilities through relations [26]

$$\begin{aligned}
\mathbf{p} &= \alpha_{ee}\mathbf{E} + \alpha_{em}\mathbf{H}, \\
\mathbf{m} &= -\mu_0^{-1}\alpha_{em}\mathbf{E} + \alpha_{mm}\mathbf{H},
\end{aligned} \quad (1)$$

where $\alpha_{ee}$ and $\alpha_{mm}$ are the electric and magnetic polarizabilities, respectively, and $\mu_0$ denotes the vacuum's magnetic permeability [see Appendix A for the definitions of dipole moments **p** and **m**]. Moreover, the magnetoelectric polarizability $\alpha_{em}$ indicates the NP's chirality, and determining this quantity is the goal in the chirality characterization. Considering that the dipole moments are not directly observable, we require to find a relation between the measurable rate of energy transfer to the particle, i.e., the transferred power, and the particle's magnetoelectric polarizability $\alpha_{em}$ [3]. The measurable quantities are absorption, scatter-ing, and extinction which are, respectively, the rates of energy dissipated, scattered, and the sum of these two energies by the particle. Considering that $u_e = \epsilon_0 |\mathbf{E}|^2 /4$ and $u_m = \mu_0 |\mathbf{H}|^2 /4$, respectively, the time-averaged electric and magnetic energy densities at the position of the NP, the time-averaged scattered power by the NP modeled by dipoles in Eq. (1) is obtained from the Poynting theorem and reads (see Appendix A)

$$\begin{aligned}
P_{sca} &= \frac{\omega k^3}{3\pi}\left(\left|\frac{\alpha_{ee}}{\epsilon_0}\right|^2 + \frac{\epsilon_0}{\mu_0}|\alpha_{em}|^2\right)u_e \\
&+ \frac{\omega k^3}{3\pi}\left(|\alpha_{mm}|^2 + c_0^2|\alpha_{em}|^2\right)u_m \\
&+ \frac{\omega k^3}{6\pi}\Re\left\{\left(\epsilon_0^{-1}\alpha_{ee}\alpha_{em}^* - \alpha_{em}\alpha_{mm}^*\right)\mathbf{H}^*\cdot\mathbf{E}\right\},
\end{aligned} \quad (2)$$

where "*" denotes complex conjugation. Moreover, $\epsilon_0$, $c_0$, and $k$ are, respectively, the permittivity, speed of light, and wavenumber in vacuum. Note that monochromatic electromagnetic fields with time dependence $\exp(-i\omega t)$ with $\omega$ being the angular frequency, are considered throughout the paper. The time-averaged extinction power is obtained as

$$P_{ext} = 2c_0^{-1}\omega\Im\{\alpha_{ee}\}u_e + 2\omega\Im\{\alpha_{mm}\}u_m + 2c_0\omega^2\Re\{\alpha_{em}\}h, \quad (3)$$

with $h = \Im\{\mathbf{E}\cdot\mathbf{H}^*\}/(2\omega c_0)$ being the time-averaged helicity density of the local field acting on the NP, discussed in detail in the next subsection since it is one of the core quantities in this work. Finally, the time-averaged absorbed power $P_{abs} = P_{ext} - P_{sca}$ is obtained from Eqs. (2) and (3). Equations (2) and (3) suggest that if we would be able to manipulate the fields in a way so that we eliminate the terms which contain polarizabilities $\alpha_{ee}$ and $\alpha_{mm}$ while simultaneously keep the terms which contain the polarizability term $\alpha_{em}$, then, we are able to characterize the particle's chirality from the corresponding measured power. However, Eq. (2) indicates that it is not a straightforward task to manipulate the local fields and eliminate the terms containing electric and magnetic polarizabilities $\alpha_{ee}$ and $\alpha_{mm}$ from the scattered and absorption powers while keeping the terms which contain the polarizability term $\alpha_{em}$. Nonetheless, the third term in the right hand side of Eq. (3), which solely depends on magnetoelectric polarizability $\alpha_{em}$ and the helicity density of the field, suggests that it is possible to dismiss the first two terms and keep the last term by properly engineering of the local fields. Therefore, we conclude that the extinction power, Eq. (3), is the most convenient candidate to reveal the chirality of NPs represented by dipole moments.

---

[3] Note that there are also other physical measurable quantities such as electromagnetic force and torque, which are out of the scope of the current study. For more information, the readers may refer to [27]–[36].

Let us next study how to dismiss the first two terms in the expression of the extinction power, Eq. (3), hence, enable chirality detection/characterization. We consider the interaction of a chiral NP with *two* distinct excitations and denote the measured extinctions in the two "experiments" by superscripts "+" and "−", respectively (the + and − signs are chosen since the two excitations possess helicities with opposite signs). The difference between the extinction powers (which we call differential extinction) in these two interactions is $\Delta P_{ext} = P_{ext}^{+} - P_{ext}^{-}$ and equals

$$\Delta P_{ext} = 2\omega\epsilon_0^{-1}\Im\{\alpha_{ee}\}\Delta u_e + 2\omega\Im\{\alpha_{mm}\}\Delta u_m + \\ 2c_0\omega^2\Re\{\alpha_{em}\}\Delta h, \tag{4}$$

where $\Delta h = \Im\{\mathbf{E}^+ \cdot (\mathbf{H}^+)^* - \mathbf{E}^- \cdot (\mathbf{H}^-)^*\}/(2\omega c_0)$, $\Delta u_e = \epsilon_0(|\mathbf{E}^+|^2 - |\mathbf{E}^-|^2)/4$, and $\Delta u_m = \mu_0(|\mathbf{H}^+|^2 - |\mathbf{H}^-|^2)/4$. It is easy to check that the first two terms in Eq. (4) containing the electric and magnetic polarizabilities are eliminated in the differential extinction $\Delta P_{ext}$ when

$$\Delta u_m = u_m^+ - u_m^- = 0, \\ \Delta u_e = u_e^+ - u_e^- = 0. \tag{5}$$

This means that the electric and magnetic local fields in the two experiments are required to be equal in magnitude, which is our suggested choice. Although conditions (5) eliminate the contributions of the electric and magnetic polarizabilities from the interactions, the remaining term under this condition, i.e.,

$$\Delta P_{ext} = 2c_0\omega^2\Re\{\alpha_{em}\}\Delta h, \tag{6}$$

contains the fields' property denoted by $\Delta h$ as well as the particle's chirality expressed by $\alpha_{em}$. Equation (6) suggests that the differential extinction $\Delta P_{ext}$ is a suitable candidate for detection of particle's chirality when $\Delta h \neq 0$. It is important to note that higher values of $\Delta h$ corresponds to a better chance for the detection of smaller values of chirality, i.e., $\Re\{\alpha_{em}\}$ according to Eq. (6). In Secs. III and IV, we extensively discuss several scenarios as the excitation fields which are capable of enhancing the possibility of chirality detection of particles.

In the next step, we determine the value of $\Re\{\alpha_{em}\}$ independent of the field properties $\Delta h$. Inspired from the works [24], [37], we consider the dissymmetry factor $g$ defined as

$$g = \frac{\Delta P_{ext}}{\bar{P}_{ext}}, \tag{7}$$

where, $\bar{P}_{ext} = (P_{ext}^+ + P_{ext}^-)/2$ is the arithmetic average of the extinction powers in the two experiments. Now, assuming $\bar{u}_e = (u_e^+ + u_e^-)/2$, $\bar{u}_m = (u_m^+ + u_m^-)/2$ and $\bar{h} = (h^+ + h^-)/2$ are, respectively, the arithmetic averages of the time-averaged electric energy, magnetic energy, and helicity densities in the two experiments, $\bar{P}_{ext}$ reads

$$\bar{P}_{ext} = 2\epsilon_0^{-1}\omega\Im\{\alpha_{ee}\}\bar{u}_e + 2\omega\Im\{\alpha_{mm}\}\bar{u}_m + 2c_0\omega^2\Re\{\alpha_{em}\}\bar{h}. \tag{8}$$

Next, by introducing the new condition

$$h^+ = -h^-, \tag{9}$$

for fields, which means that fields in the two experiments possess helicity densities with equal amplitudes and opposite signs and implies higher possibilities of detection of NPs' chirality for fields satisfying conditions (5) (see Ref. [1] for discussion), the dissymmetry factor $g$ reduces to

$$g = \frac{2c_0\omega\Re\{\alpha_{em}\}h^+}{\epsilon_0^{-1}\Im\{\alpha_{ee}\}u_e^+ + \Im\{\alpha_{mm}\}u_m^+}. \tag{10}$$

Now, following Ref. [1] we introduce the convenient conditions

$$\begin{cases} |\mathbf{E}^\pm| = \eta_0|\mathbf{H}^\pm|, \\ \Im\{\mathbf{E}^\pm \cdot (\mathbf{H}^\pm)^*\} = \pm|\mathbf{E}^\pm||\mathbf{H}^\pm|. \end{cases} \tag{11}$$

for each of the two excitations, where $\eta_0$ is the intrinsic impedance of vacuum, and call the fields satisfying these condition *optimally chiral*. Conditions (11) result in two important consequences: one, the time-averaged helicity and energy densities in the two experiments follow the relations $|h^\pm| = u^\pm/\omega$ since $u^\pm = u_e^\pm + u_m^\pm = 2u_e^\pm = 2u_m^\pm$ and; two, the helicity density $|h^\pm| = u^\pm/\omega$ is the maximum achievable helicity density for a given energy density (see Ref. [1] for a complete discussion). Therefore, the dissymmetry factor $g$ for those fields that satisfy conditions (5), (9), and (11) reads

$$g = \frac{4c_0 \Re\{\alpha_{em}\}}{\epsilon_0^{-1}\Im\{\alpha_{ee}\} + \Im\{\alpha_{mm}\}}, \tag{12}$$

and it measures the particle's chirality with respect to its electric and magnetic polarizabilities independent of the field's property which was denoted by the field's helicity and energy densities. Moreover, for particles with weak magnetic polarizabilities, i.e., $\epsilon_0^{-1}\Im\{\alpha_{ee}\} >> \Im\{\alpha_{mm}\}$, the dissymmetry factor $g$ reduces to

$$g = 4\eta_0^{-1}\Re\{\alpha_{em}\} / \Im\{\alpha_{ee}\} \tag{13}$$

and measures the ratio between the magnetoelectric and electric polarizability, i.e., it characterizes the NP's chirality assuming that the NP's electric polarizability $\alpha_{ee}$ is known.

### B. Helicity in chiral fields

In this subsection, the helicity of fields, which is a quantitative measure of the fields' curliness and plays an important role in the chirality detection of particles, is briefly reviewed and discussed.

Conservation of physical quantities such as energy and momentum along with the Maxwell's equations enable us to define some properties such as linear momentum, angular momentum, chirality, etc. for electromagnetic fields analogous to the mechanical and geometrical properties for matter [25]. One of such properties which is related to the conservation of spin angular momentum, is helicity[38]. The helicity of electromagnetic fields which, after time-averaging, reads

$$H = \int h\, d\boldsymbol{u} = \frac{1}{4c_0}\int\left(\epsilon_0 \mathbf{F}^* \cdot \nabla \times \mathbf{F} + \mu_0^{-1}\mathbf{A}^* \cdot \nabla \times \mathbf{A}\right)d\boldsymbol{u}, \tag{14}$$

was introduced to show how lines of field vectors are *linked* [39], [40]. Note that the integration is carried out over the entire space, and $\mathbf{F}$ and $\mathbf{A}$ are, respectively, the electric and the magnetic vector potentials [41]–[43]. To perceive the physical significance of the helicity in Eq. (14), one expands an electromagnetic field into its spectral components with right and left handed circular polarization. This illustrates that the total helicity is the average of helicities of all photons, defined as the projection of their angular momenta along their linear momenta[44], and indeed it is the classical limit for the difference between numbers of photons with right-handed and left-handed circular polarizations [42] and represents the fields' handedness. Moreover, investigating the interaction of particles with electromagnetic fields as discussed earlier in this section demonstrates the significance of time-averaged helicity density of fields, i.e.,

$$h = \frac{1}{4c_0}\left(\epsilon_0 \mathbf{F}^* \cdot \nabla \times \mathbf{F} + \mu_0^{-1}\mathbf{A}^* \cdot \nabla \times \mathbf{A}\right) \tag{15}$$

in the chirality detection of particles. Note that the time-averaged helicity density for monochromatic fields is proportional to their time-averaged *optical chirality*[37]

$$c = \frac{1}{4}\left(\epsilon_0 \mathbf{E}^* \cdot \nabla \times \mathbf{E} + \mu_0^{-1}\mathbf{B}^* \cdot \nabla \times \mathbf{B}\right) \tag{16}$$

through the relation $c = \omega^{-2}c_0^{-1} h$. However, optical chirality is related to the conservation of the spin angular momentum associated with the curl of fields [38] (versus the standard angular momentum associated to the electromagnetic fields). Therefore, we adopt the helicity as a measure of fields' capability in the detection of particles' chirality and refer to the fields with nonzero helicity density as *chiral fields*.

## III. CHIRAL OPTICAL BEAMS

As mentioned earlier, in this section we discuss engineered beams that enhance the possibility of chirality detection and enable characterization of NPs' chirality. Indeed, we investigate the helicity densities of two chiral optical beams and demonstrate how they offer the possibility of enhancing helicity densities by manipulating their field profiles and polarizations.

### A. Gaussian beams with circular polarization

We first consider a Gaussian beam (GB) with circular polarization. Suppressing the time dependence $\exp(-i\omega t)$, the electric field vector of a GB which travels in the $+z$ direction under the paraxial approximation reads [45]

$$\mathbf{E} = \frac{V_G}{w}e^{-(\rho/w)^2}e^{ik\rho^2/(2R)}e^{-i\tan^{-1}(z/z_R)}e^{ikz}(\hat{\mathbf{x}} \pm i\hat{\mathbf{y}}), \tag{17}$$

where $+/-$ sign, represents right/left-handed circular polarizations, and $\hat{\mathbf{x}}$ and $\hat{\mathbf{y}}$ are the unit vectors in the Cartesian coordinate system. Here, $z_R = \pi w_0^2 / \lambda$, $w = w_0\sqrt{1+(z/z_R)^2}$, and $R = z[1+(z_R/z)^2]$ with $w_0$ and $\lambda$ being, respectively, the beam parameter and the excitation wavelength. The beam parameter represents half a beam waist when the beam is not tightly focused[46]. Moreover, $\rho$ and $V_G$ are, respectively, the radial distance from the beam axis (here the $z$-axis of the Cartesian coordinate system) and the complex amplitude of the beam (with the unit of Volt). The magnitude of the electric field of a typical GB with $w_0 = \lambda$ in the $x$-$z$ plane and the electric field vectors along the propagation direction (here the $z$-axis) are depicted in FIG. 1 (a) and (b), respectively.

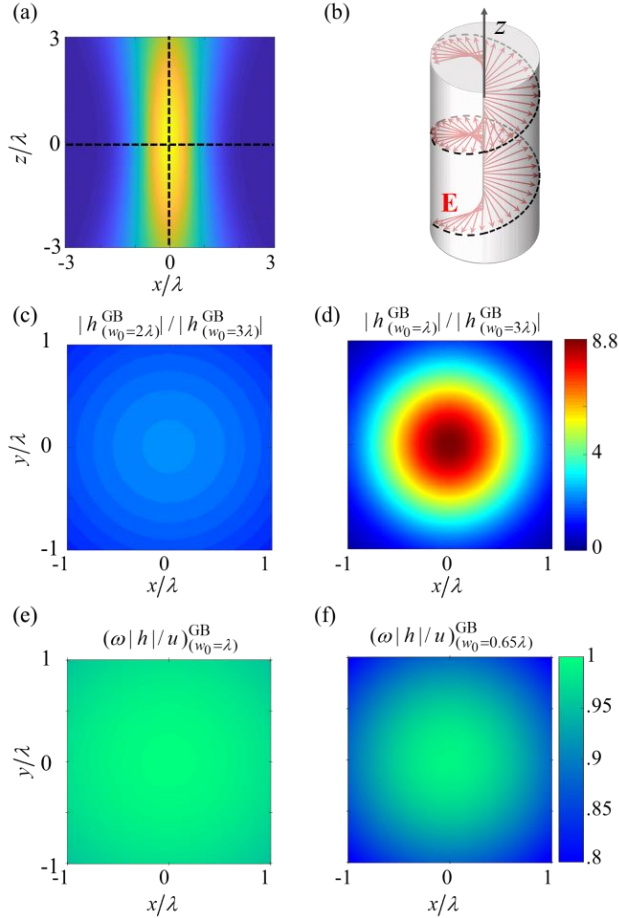

FIG. 1. (a) Electric field magnitude in the $x$-$z$ plane and, (b) propagation of the electric field vector along the $z$-direction for a Gaussian beam with circular polarization and $w_0 = \lambda$ propagating along the $z$ direction. Helicity density $h$ of the GB on the $z = 0$ transverse plane with (c) $w_0 = 2\lambda$, and (d) $w_0 = \lambda$, normalized to that of a GB with the same power and $w_0 = 3\lambda$. Note that the helicity density at the center of the coordinate system of the beam with $w_0 = \lambda$ is approximately four times stronger than that of the beam with $w_0 = 2\lambda$, confirming that the helicity density is enhanced when the beam parameter $w_0$ decreases i.e., for tighter focused beams. The quantity $\omega |h|/u$ for circularly polarized GB with (e) $w_0 = 0.65\lambda$ and (f) $w_0 = \lambda$.

Next, we illustrate the helicity densities of two circularly polarized GBs with equal total power and different beam parameters $w_0 = \lambda$ and $w_0 = 2\lambda$ normalized to the helicity density of a GB with the same power and $w_0 = 3\lambda$, all evaluated at the $z = 0$ plane in FIG. 1 (c) and (d), respectively. We recall that by increasing the beam parameter the GB looks more like a plane wave. As it is clear from these figures, reducing the beam parameter $w_0$ results in an enhancement in the helicity density of the beam around its optical axis. This is due to the localization of the energy density at these points which, based on our previous discussions, improves the possibility of chirality detection.

It is noteworthy that the GB in Eq. (17) is obtained under paraxial approximation and it optimally chiral, i.e., satisfies conditions (11), only when $\rho \ll w_0$ [see Appendix C], which is a restricted area in space, since under these conditions the fields of a GB look more like those of a plane wave. The quantity $\omega |h|/u$, which takes the value of 1 when the conditions (11) are satisfied, is depicted in FIG. 1 (e) and (f) for circularly polarized GBs with different values of beam parameters. The deviation of this quantity from unity by moving away from the beam axis means that tightly focused GBs are not the best candidate for chirality characterization since conditions (11) are not precisely satisfied everywhere in space. Therefore, in the next step we propose to use an engineered structured light beam with the interesting property of satisfying conditions (11) everywhere in space.

### B. Superposition of azimuthally and radially polarized beams

A GB with circular polarization is a trivial optimally chiral beam. However, it would be interesting if we could introduce an alternative optimally chiral optical beam with different properties). Here we propose the superposition of an Azimuthally Polarized Beam (APB) and a Radially Polarized Beam (RPB), which we call ARPB, as a chiral optical beam that can be used to characterize the chirality of NPs. The APB and ARB are obtained by superposing Laguerre Gaussian beams with opposite angular momenta. Let us study the distinctive features of the ARPB compared to a GB with circular polarization. The electric field vector in an APB which is propagating along the $z$-direction (optical axis of the beam) is polarized along the azimuthal direction $\hat{\boldsymbol{\varphi}}$ in the transverse plane and reads [46]

$$\mathbf{E}^{\mathrm{APB}} = \frac{2V_A \rho}{\sqrt{\pi} w^2} e^{-(\rho/w)^2 \zeta} e^{-2i \tan^{-1}(z/z_R)} e^{ikz} \hat{\boldsymbol{\varphi}}, \qquad (18)$$

where $V_A$ (with the unit of Volt) is the complex amplitude of the beam and $\zeta = 1 - iz/z_R$. Other parameters are similar to those defined for a GB in Eq. (17). It is important to note that such a beam has a zero electric field and a nonzero longitudinal magnetic field component along its optical axis. In duality with APBs, RPBs have magnetic field vectors which are polarized along the azimuthal direction and read

$$\mathbf{H}^{\mathrm{RPB}} = \frac{2V_R \rho}{\eta_0 \sqrt{\pi} w^2} e^{-(\rho/w)^2 \zeta} e^{-2i \tan^{-1}(z/z_R)} e^{ikz} \hat{\boldsymbol{\varphi}}, \qquad (19)$$

with a complex amplitude $V_R$ (with the unit of Volt). Note, an RPB has a nonzero longitudinal electric field component along its axis (for the longitudinal field expressions see Refs.

[34], [45], [46]).We propose the superposition of these two vortex beams, i.e.,

$$\mathbf{E}^{\text{ARPB}} = \mathbf{E}^{\text{APB}} + \mathbf{E}^{\text{RPB}},$$
$$\mathbf{H}^{\text{ARPB}} = \mathbf{H}^{\text{APB}} + \mathbf{H}^{\text{RPB}}, \qquad (20)$$

with the same beam parameters and respective amplitudes such that $V_A = \pm i V_R^*$. Indeed, by applying the source-free Maxwell's curl equations $\nabla \times \mathbf{E} = i\omega\mu_0\mathbf{H}$ and $\nabla \times \mathbf{H} = -i\omega\varepsilon_0\mathbf{E}$ it is easy to deduce that an ARPB possesses both the electric and magnetic field components along the azimuth direction $\hat{\boldsymbol{\phi}}$ as well as along the radial and longitudinal directions $\hat{\boldsymbol{\rho}}$ and $\hat{\mathbf{z}}$. Under the above situation the proposed ARPB combination constitutes an optimally chiral chiral optical beam that satisfies conditions (11) everywhere in space. FIG. 2 (a) and (b) demonstrate magnitude of the electric fields in $x$-$z$ and $x$-$y$ planes, respectively. Note that the field magnitude reaches its maximum at $z = 0$ where $w = w_0$.

There are two important distinctions between the ARPB and a GB with circular polarization. First, the electric and magnetic field vectors in an ARPB are parallel, everywhere, with a phase difference of $\pi/2$ at all points in space when choosing $V_A = \pm i V_R^*$ in contrast to a GB where a circular polarization is obtainable only in a limited area around the beam axis. Second, the ARPB has field vectors at the beam axis which are exclusively longitudinal. This provides the opportunity of characterization of longitudinal chirality in contrast to a GB with circular polarization that supports the characterization of a transverse chirality [34]. Figures 2 (c) and (d) depict the helicity densities of two ARPBs with $V_A = \pm i V_R^*$ and beam waist parameter values of $w_0 = \lambda$ and $w_0 = 0.65\lambda$, respectively, in the plane $z = 0$ normalized to the helicity density of a circularly polarized GB with $w_0 = 3\lambda$ and the same power. As it is clear, this structured light satisfying conditions (11) results in the helicity density enhancement and such an enhancement will in turn improves the possibility of chirality detection of particles as discussed in Sec. II. As a final note of this subsection, by comparing FIG. 2 (c) and (d) with FIG. 1 (c) and (d) one observes another distinction between the ARPB and the GB with circular polarization. That is, a GB with circular polarization reaches its maximum helicity density at the beam axis whereas an ARPB attains its maximum helicity density on an annular ring around the beam axis.

So far, we have studied how to enhance the helicity density by engineering an optical beam and we have achieved about an order of magnitude enhancement [see FIG. 2 (d)]. In the next section, however, we propose an alternative route to further enhance helicity density.

## IV. CHIRAL NEARFIELD AND NANOANTENNAS

We propose the use of specific nano-antennas (NAs) to enhance helicity density. We show how the nearfield of a properly engineered NA, excited by a chiral beam, provides at least one extra order of magnitude helicity density enhancement compared to the helicity of the exciting beam. The total nearfield of a NA includes the contributions from both the incident optical beam and scattered nearfields, therefore, its helicity density $h$ reads

$$h = h_{\text{inc}} + h_{\text{sca}} + h_{\text{int}}, \qquad (21)$$

where $h_{\text{sca}} = \Im\{\mathbf{E}_{\text{sca}} \cdot \mathbf{H}_{\text{sca}}^*\}/(2\omega c_0)$ and $h_{\text{inc}} = \Im\{\mathbf{E}_{\text{inc}} \cdot \mathbf{H}_{\text{inc}}^*\}/(2\omega c_0)$ are, respectively, the helicity densities of the scattered and incident fields. Moreover, $h_{\text{int}}$, which we call interference helicity density, reads

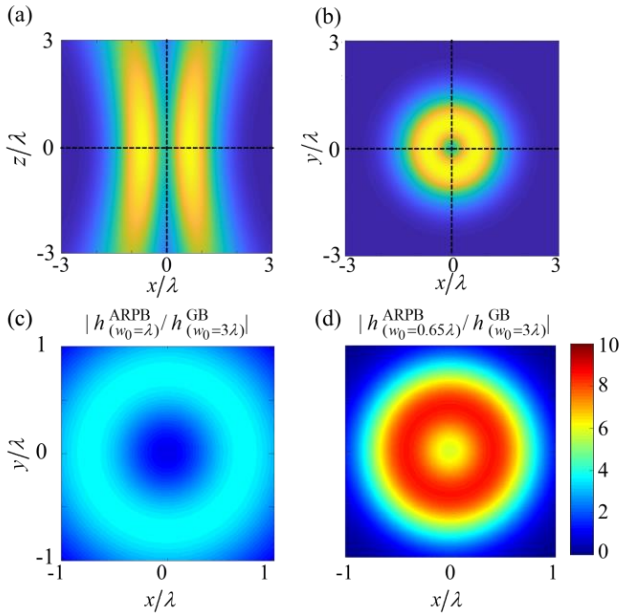

FIG. 2. A chiral ARPB propagating along the $z$-direction with the beam waist parameter $w_0 = \lambda$. Electric field magnitude in the $x$-$z$ plane (a), and in the $x$-$y$ plane (b). Helicity density of the ARPB in the $x$-$y$ plane normalized to that of a standard circularly polarized GB with a same power and $w_0 = 3\lambda$ : in (c) the APRB has $w_0 = \lambda$, whereas in (d) $w_0 = 0.65\lambda$. Note that the helicity density of the ARPB with $w_0 = 0.65\lambda$ on the beam axis, is 6 times stronger than that of an ARPB with $w_0 = \lambda$.

$$h_{\text{int}} = \frac{1}{2\omega c_0} \Im \left\{ \mathbf{E}_{\text{sca}} \cdot \mathbf{H}_{\text{inc}}^* + \mathbf{E}_{\text{inc}} \cdot \mathbf{H}_{\text{sca}}^* \right\}. \qquad (22)$$

Equation (21) suggests that for an improvement of the helicity density we need to simultaneously boost all the helicity densities $h_{\text{sca}}$, $h_{\text{inc}}$, and $h_{\text{int}}$. To elaborate further, we illuminate the NA by chiral beams introduced in the previous section and devote the rest of this section to the relation between the helicity density of the scattered nearfield $h_{\text{sca}}$, the properties of the required NA for the optimum performance, and the influential factors on the interference helicity density $h_{\text{int}}$.

### A. Concepts of nanoantennas for helicity enhancement

Let us assume that a NA is located at the center of Cartesian coordinates and it is illuminated by a chiral beam propagating along the $+z$ direction (FIG. 3).

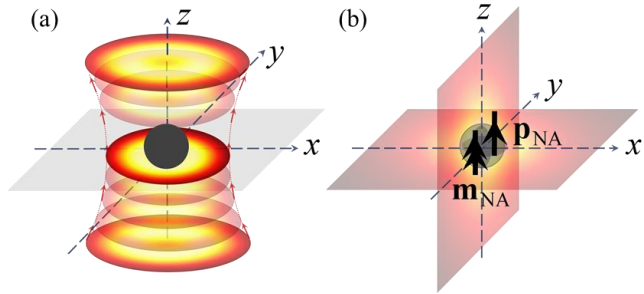

FIG. 3. The nearfield of a NA with an enhanced helicity density with respect to that of an incident chiral beam. (a) An optical beam propagating along the $z$ direction illuminates the proposed NA which is located at the center of the coordinate system. (b) The proposed NA is modeled by induced electric and magnetic dipole moments $\mathbf{p}_{\text{NA}}$ and $\mathbf{m}_{\text{NA}}$, respectively, and their scattered nearfields demonstrate high helicity density in the close vicinity of the NA.

With the goal of investigating the helicity density of the NA and without loss of generality, we analyze an isotropic NA (a nanosphere) operating in its dipolar regime, and model it by the induced electric and magnetic dipole moments $\mathbf{p}_{\text{NA}}$ and $\mathbf{m}_{\text{NA}}$, respectively, which are located at the origin of the Coordinate system. Following the discussions in Appendix B, the dominant contribution of the helicity density $h_{\text{sca}}$ of scattered fields in the near-zone of the NA at the radial positions $\mathbf{r} = r\hat{\mathbf{r}}$ reads

$$h_{\text{sca}} \approx \frac{\eta_0}{32\pi^2 \omega r^6} \Im \left\{ 3(\hat{\mathbf{r}} \cdot \mathbf{p}_{\text{NA}})(\hat{\mathbf{r}} \cdot \mathbf{m}_{\text{NA}}^*) + \mathbf{p}_{\text{NA}} \cdot \mathbf{m}_{\text{NA}}^* \right\}. \qquad (23)$$

Note that in evaluating Eq. (23) in the nearfield of the NA, we only consider the dominant terms with $r^{-3}$ dependence [see Eqs. (B1) and (B2) in Appendix B]. The term $\Im \{\mathbf{p}_{\text{NA}} \cdot \mathbf{m}_{\text{NA}}^*\}$ in Eq. (23) not only relates helicity to the strength of induced electric and magnetic dipoles of the NA but also implies that the dipoles relative spatial orientation and phase should be controlled to maximize helicity density of the scattered nearfields. Note that at location vectors $\mathbf{r}$ that are parallel to both dipole moments, the term $\Im \{3(\hat{\mathbf{r}} \cdot \mathbf{p}_{\text{NA}})(\hat{\mathbf{r}} \cdot \mathbf{m}_{\text{NA}}^*)\}$ has a constructive contribution to the enhancement of the helicity density $h_{\text{sca}}$. In other words, these location vectors define the regions where helicity around a NA is the strongest. The dipole moments $\mathbf{p}_{\text{NA}}$ and $\mathbf{m}_{\text{NA}}$ are, respectively, related to the incident electric and magnetic fields $\mathbf{E}_{\text{inc}}^o$ and $\mathbf{H}_{\text{inc}}^o$ at the position of the NA (superscript "o" denotes the NA's position which is the origin of the Coordinate system in our example) through the electric and magnetic polarizabilities $\alpha_{\text{ee}}^{\text{NA}}$ and $\alpha_{\text{mm}}^{\text{NA}}$ of the NA as

$$\mathbf{p}_{\text{NA}} = \alpha_{\text{ee}}^{\text{NA}} \mathbf{E}_{\text{inc}}^o,$$
$$\mathbf{m}_{\text{NA}} = \alpha_{\text{mm}}^{\text{NA}} \mathbf{H}_{\text{inc}}^o. \qquad (24)$$

Since we consider optimally chiral incident fields, i.e., $\mathbf{E}_{\text{inc}}^o = \pm i\eta_0 \mathbf{H}_{\text{inc}}^o$, and when $\mathbf{m}_{\text{NA}} = c_0 \mathbf{p}_{\text{NA}}$ the scattered nearfield of the proposed NA is also optimally chiral [see Appendix B], then, Eq. (24) implies the balance relation

$$\alpha_{\text{ee}}^{\text{NA}} = \varepsilon_0 \alpha_{\text{mm}}^{\text{NA}} \qquad (25)$$

between the polarizabilities of the NA. Note that Eq. (25) ensures the relation $\mathbf{E}_{\text{sca}} = \pm i\eta_0 \mathbf{H}_{\text{sca}}$ holds everywhere in space and not only in the near zone of the NA, indicating that condition (25) implies the best possible scattered nearfields everywhere. Therefore, the problem of obtaining the maximum achievable helicity density of the scattered nearfields is reduced to designing a NA that satisfies the balance relation (25). Such a balance relation proposes that we require a NA which simultaneously possesses both electric and magnetic polarizabilities. Though materials with significant magnetic properties are not available at optical frequencies [47], "resonant magnetism" is possible for example with plasmonic clusters [48]–[56] or dielectric nanostructures [57]–[69] that supports Mie resonances. An optimum design with a high helicity density (several orders of magnitude) requires a separate study, however, in the next step and only with the goal of a proof of concept, we propose a very simple NA design with spherical shape which is made of a high index material and demonstrate an enhancement higher than an order of magnitude in the helicity density with respect to the excitation beam.

To that end, we consider a spherical silicon (Si) nanoparticle with radius $a$ and plot the normalized (to its maximum)

quantity $|\alpha_{ee}^{NA} - \epsilon_0 \alpha_{mm}^{NA}|$ versus wavelength $\lambda$ and radius $a$ , in a logarithmic scale, in FIG. 4. The quantity $|\alpha_{ee}^{NA} - \epsilon_0 \alpha_{mm}^{NA}|$ vanishes when the balance relation (25) is satisfied. Note that for an efficient NA design, we require materials with low losses. Since for wavelengths smaller than 800 nm the material loss in amorphous Si is not negligible we employ crystalline Si for this wavelength range. In the following discussion on Si NA, we perform our calculations in two frequency ranges and in each range with either crystalline or amorphous Si: for wavelength range from 500 nm to 800 nm we employ crystalline Si, and for wavelength range from 800 nm to 1200 nm we employ amorphous Si.

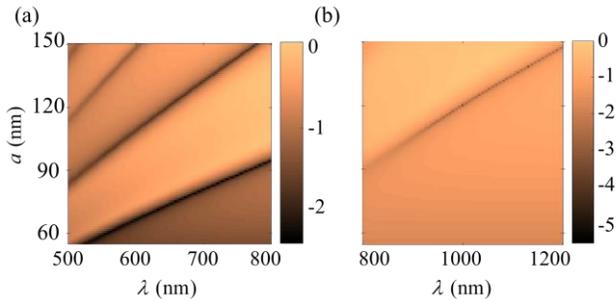

FIG. 4. The normalized (to its maximum ) values of $|\alpha_{ee}^{NA} - \epsilon_0 \alpha_{mm}^{NA}|$ , in logarithmic scale, versus the radius $a$ of the NA and wavelength $\lambda$ of the excitation field for (a) crystalline silicon[70], and (b) amorphous silicon[71]. Note that large negative values (dark regions) indicate that the balance relation $\alpha_{ee}^{NA} = \epsilon_0 \alpha_{mm}^{NA}$ is satisfied. The polarizabilities are calculated using the Mie scattering coefficients[72].

Note that when condition (25) holds for the NA (e.g. on the dark regions in FIG. 4), the scattered nearfield, and consequently, the total fields is optimally chiral. Moreover, although the spatial distribution of the helicity density of the scattered nearfield is non-uniform around the NA, the relation $/h/= u/\omega$ holds everywhere in space as long as $\alpha_{ee}^{NA} = \epsilon_0 \alpha_{mm}^{NA}$ is valid for the proposed NA.

It is now appropriate to discuss practical scenarios when the engineered NA is exposed to external illuminations and investigate the enhancement of the helicity density. We first consider the case when the NA is illuminated with an ARPB since its analytical investigation is more straightforward and then consider the illumination with a circularly polarized GB. Finally, we note that full-wave analysis shows that the quadrupoles for the considered parameters $\lambda$ and $a$ are negligible since $a/\lambda \ll 1$.

### B. ARPB illumination

We consider a situation when the engineered NA is illuminated by an ARPB propagating in the $+z$ direction with $w_0 = \lambda$ , where the electric and magnetic field vectors are oriented along the $z$ direction, at the position of the NA, as discussed in Sec. IIIB. Since the NA is azimuthally symmetric, the induced electric and magnetic dipole moments $\mathbf{p}_{NA}$ and $\mathbf{m}_{NA}$ are parallel to the $z$-axis under such an illumination, and the helicity density of the scattered nearfield, introduced in Eq. (23), at location $(r, \theta, \phi)$ near the NA reads

$$h_{sca} \approx \frac{|\mathbf{E}_{inc}^{o}|^2}{32\pi^2 \omega r^6} \left(3\cos^2\theta + 1\right) \Re\left\{\alpha_{ee}^{NA} \left(\alpha_{mm}^{NA}\right)^*\right\}. \tag{26}$$

Now we define the enhancement factor $/h_{sca}/ / /h_{inc}^{o}/$ , which is the ratio of scattered nearfield helicity density $h_{sca}$ to the helicity density of the incident field at the origin where the NA is located. Note that the selection of helicity density of the incident field at the position of the NA, $h_{inc}^{o}$ , eliminates the influence of the incident field intensity from the enhancement factor $/h_{sca}/ / h_{inc}^{o}/$ and consequently this enhancement shows the helicity enhancement due to the scattered nearfields of the NA. Since the electric $\mathbf{E}_{inc}$ and magnetic $\mathbf{H}_{inc}$ components of the incident field satisfy $\mathbf{E}_{inc} = \pm i\eta_0 \mathbf{H}_{inc}$ , the helicity density of the incident field at the position of the NA reads $/h_{inc}^{o}/ = |\mathbf{E}_{inc}^{o}\| \mathbf{H}_{inc}^{o}| /(2\omega c_0)$ . Consequently, the helicity enhancement due to scattered nearfields of the NA is approximated as

$$\frac{|h_{sca}|}{|h_{inc}^{o}|} \approx \frac{1}{16\epsilon_0 \pi^2 r^6} \left|\left(3\cos^2\theta + 1\right) \Re\left\{\alpha_{ee}^{NA} \left(\alpha_{mm}^{NA}\right)^*\right\}\right|. \tag{27}$$

It is clear from Eq. (27) that the maximum helicity density occurs for $\theta = 0, \pi$ , i.e., along the $z$ direction. We also define the energy enhancement ratio $u_{sca} / u_{inc}^{o}$ as the time-averaged energy density of the scattered nearfield at a desired location with respect to the incident energy density at the position of the NA. The helicity and energy density enhancements $/h_{sca} / h_{inc}^{o}/$ and $u_{sca} / u_{inc}^{o}$ for various values of radius $a$ of the spherical NA versus different wavelength $\lambda$ , evaluated at the surface of the NA along the $z$ direction, i.e., $\mathbf{r} = a\hat{\mathbf{z}}$ , for nanospheres with both crystalline and amorphous Si are illustrated in FIG. 5. Note that in Eq. (27) we only considered the dominant terms of the nearfield of the NA, generated by the superposition of an electric and a magnetic dipole, to provide a simple analytical formula. However the values demonstrated in FIG. 5 include all the term of the spherical Greens function for completeness, and we note that these values are very close to those provided by the approximate formula given in Eq. (27).

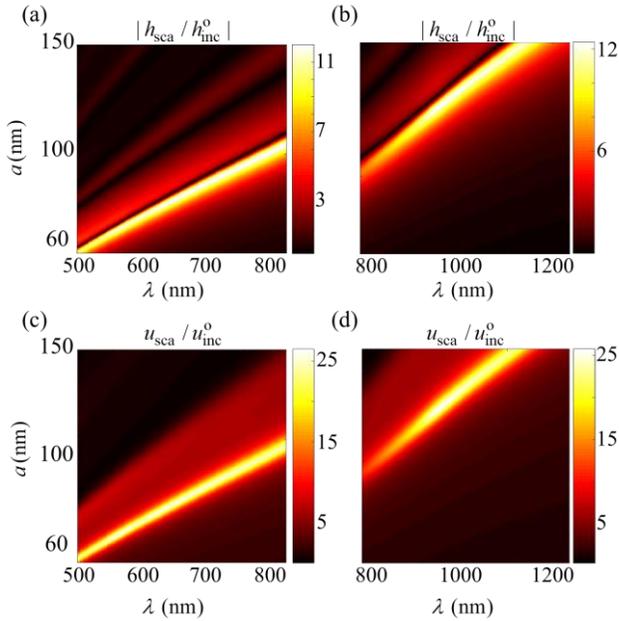

FIG. 5. Enhancement of helicity and energy densities on the surface of the proposed NA, evaluated along the $z$ direction, i.e., at $\mathbf{r} = a\hat{\mathbf{z}}$, when illuminated by an ARPB propagating along the $z$ direction with the beam parameter $w_0 = \lambda$. Helicity enhancement in the scattered nearfield is evaluated for (a) crystalline and (b) amorphous silicon NA. Energy enhancement in the scattered nearfield is evaluated for (c) crystalline and (d) amorphous silicon NA.

Next, using the evaluated helicity and energy densities at the boundary of the NA, i.e., where $\mathbf{r} = a\hat{\mathbf{z}}$, in FIG. 6 we plot the required radius of the NA versus excitation wavelength $\lambda$ to enforce: (a) that the balanced relation (25) is satisfied [which is equivalent to having scattered nearfield satisfying the condition $|h_{sca}| = u_{sca}/\omega$]; (b) that the maximum helicity enhancement $h_{max}$ is achieved; and (c) that the maximum energy density enhancement $u_{max}$ is achieved. In all the cases the power of the incident field is kept constant.

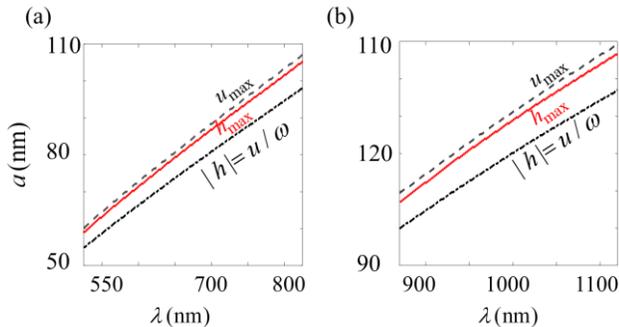

FIG. 6. Radius of the NA that guarantees the maximum enhancement of the energy $u_{max}$ and the helicity $h_{max}$ densities, and that the condition $|h| = u/\omega$ to be satisfied, versus free space wavelength $\lambda$ for (a) crystalline and (b) amorphous silicon NAs. In all cases the power of the incident field is kept constant. Note that the maximum of the helicity and the energy enhancements and the condition $\alpha_{ee}^{NA} = \epsilon_0 \alpha_{mm}^{NA}$ occur at different radius values of the Si NA, at each chosen operating wavelength. The maximum helicity enhancement occurs at radial values between the other two.

As it is clear from this figure, the maximum helicity and energy densities $h_{max}$ and $u_{max}$ do not occur for same NA radii choices. Indeed, the concept of helicity maximization indicates the maximum helicity density of the field at a fixed energy density. This means that the field may reach higher helicity density either by increasing its energy density or getting closer to the conditions (11). In the case of spherical dielectric NA, at the radius where the maximum of energy density enhancement occurs the conditions (11) are not precisely satisfied, which means that although the energy density of the field is enhanced considerably, the helicity density does not reach its upper bound $|h| = u/\omega$. Consequently, the maximum of helicity density enhancement $h_{max}$ curve locates between the curve of $u_{max}$ and that corresponding to the condition $|h| = u/\omega$. Moreover, the curve corresponding to the required condition $|h| = u/\omega$ which is essential for the characterization of chirality of particles does not cross the curve associated to $h_{max}$. Therefore, if the goal is to characterize the chirality of a NP, it is impossible to take a full benefit from the maximum achievable helicity with a dielectric nanosphere as a NA. However, if the goal is to only detect whether the particle is chiral, we can relax the condition $|h| = u/\omega$ to take a full advantage of maximum achievable helicity density. In any case, whether the detection is our goal or the characterization of the chiral polarizability $\alpha_{em}$ of a NP [see Eqs. (6) and (12)] there is always an enhancement of at least an order of magnitude in the helicity density of the scattered nearfield when employing a properly engineered Si sphere as the NA [see e.g. FIG. 8 (d)].

So far, we have discussed about the helicity enhancement due to the scattering nearfields of the NA. However, the overall helicity density of the field around the particle is determined by Eq. (21) which indicates that the overall helicity density sensed by the particle contains three contributions; that of the incident and scattered nearfields as well as that of the interference between these two helicity densities. Therefore, in the following, we discuss the importance of the interference term $h_{int}$ [see Eq. (22)] with mixed excitation and scattered nearfield terms in the helicity enhancement.

Under the balance condition (25) for NAs and considering the optimally chiral ARPB excitation, the interference helicity density given in Eq. (22) reduces to

$$h_{\text{int}} = \frac{1}{4\pi\omega r^3} \Re\left\{ e^{ikr}\alpha_{\text{ee}}^{\text{NA}} \left[ 3\left(\hat{\mathbf{r}} \cdot \mathbf{E}_{\text{inc}}^*\right)\left(\hat{\mathbf{r}} \cdot \mathbf{E}_{\text{inc}}^{\text{o}}\right) - \mathbf{E}_{\text{inc}}^{\text{o}} \cdot \mathbf{E}_{\text{inc}}^* \right] \right\}.$$
(28)

Enhancement of both interference helicity density $/h_{\text{int}}/h_{\text{inc}}^{\text{o}}$ and total helicity density $/h/h_{\text{inc}}^{\text{o}}$, evaluated at the surface of the NA along the $+z$ direction, i.e., at $\mathbf{r} = a\hat{\mathbf{z}}$, is illustrated in FIG. 7 for the two cases, i.e., for nanospheres with both the crystalline and amorphous silicon materials. As it is obvious from this figure, although the enhancement contribution due to interference helicity $h_{\text{int}}$ is weaker than that associated to the scattering fields, its contribution is not negligible in the total helicity enhancement.

Finally, three illustrative NA designs, considering the operational wavelength of $\lambda = 680\,\text{nm}$, are described in FIG. 8, with three NA radii of $a = 85$, 84, and 78 nm. These three values are chosen because they, respectively, generate: (a) the maximum enhanced energy density $u_{\text{max}}/u_{\text{inc}}^{\text{o}}$, (b) the maximum enhanced total helicity density $/h/h_{\text{inc}}^{\text{o}}$, and (c) the optimum condition $/h/= u/\omega$ for chirality characterization. The figure shows the distribution of the helicity and energy densities around the Si NA. As it is clear, an enhancement of 20-fold in helicity density which is localized at $\mathbf{r} = a\hat{\mathbf{z}}$ (in $+z$ direction) is achieved with a spherical Si NA with a radius of $a = 84\,\text{nm}$. It is noteworthy to recall that employing $V_{\text{A}} = iV_{\text{R}}^*$ and $V_{\text{A}} = -iV_{\text{R}}^*$ to measure $P_{\text{ext}}^+$ and $P_{\text{ext}}^-$ when illuminating the NA with an ARPB guarantees the satisfaction of conditions (5) and (9).

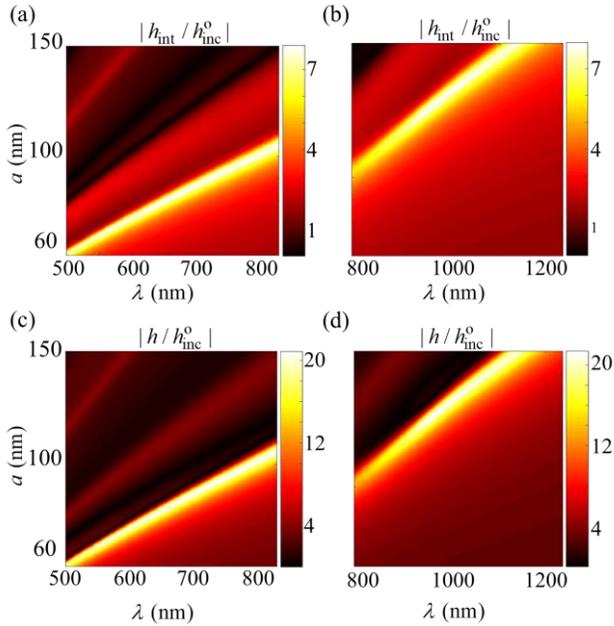

FIG. 7. The interference helicity has a considerable contribution to the total helicity enhancement. Interference helicity enhancement for (a) crystalline (b) amorphous silicon NA. Total helicity enhancement for (c) crystalline (d) amorphous silicon NA.

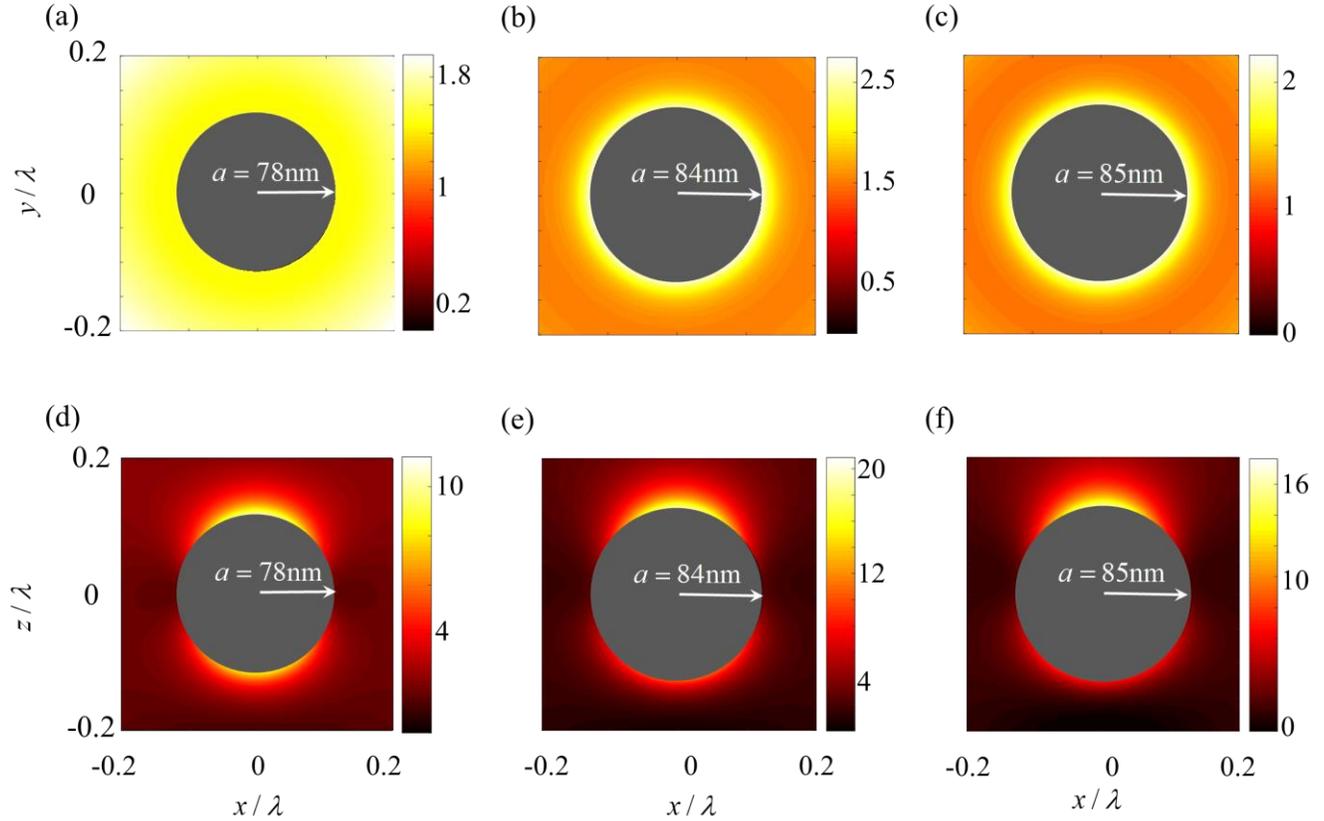

FIG. 8. The helicity enhancement $|h/h_{inc}^o|$ around the NA under ARPB illumination with $w_0 = \lambda$ in (a-c) the $x$-$y$ and (d-f) the $x$-$z$ planes at $\lambda = 680\,\text{nm}$. For the NA with radius $a = 78\,\text{nm}$ the conditions (11) are satisfied while for $a$=84 and 85 nm we have the maximum helicity and maximum energy enhancements, respectively.

### C. Gaussian excitation

To demonstrate the applicability of the proposed spherical NA with high dielectric constant to other illumination scenarios, we examine the helicity enhancement by the proposed NA when illuminated by a Gaussian beam with circular polarization. In contrast to ARPB, the GB satisfies conditions (11) only in limited areas in space as discussed in section III, in the case of tightly focused beams. Therefore, to satisfy conditions (11) in the nearfield of the NA, we set the beam parameter to $w_0 = 3\lambda$. In this scenario, the induced dipole moments $\mathbf{p}_{NA}$ and $\mathbf{m}_{NA}$ have both $x$ and $y$ components since the electric and magnetic field vectors of the illumination beam are in the $x$-$y$ plane. To ensure that the scattered nearfields is optimally chiral the induced dipole moments are required to satisfy relations $m_{NA,x} = -c_0 p_{NA,y}$, $m_{NA,y} = c_0 p_{NA,x}$, and $p_{NA,y} = i p_{NA,x}$. This results in the helicity density of the scattered nearfield to take the form

$$h_{sca} = \frac{|\mathbf{E}_{inc}^o|^2}{32\pi^2 \omega r^6} \left(\frac{3}{2}\sin^2\theta + 1\right)\Re\left\{\alpha_{ee}^{NA}\left(\alpha_{mm}^{NA}\right)^*\right\}. \quad (29)$$

Hence, the helicity enhancement $|h_{sca}/h_{inc}^o|$ reads

$$\frac{|h_{sca}|}{|h_{inc}^o|} \approx \frac{1}{16\epsilon_0\pi^2 r^6}\left|\left(\frac{3}{2}\sin^2\theta + 1\right)\Re\left\{\alpha_{ee}^{NA}\left(\alpha_{mm}^{NA}\right)^*\right\}\right|, \quad (30)$$

and the interference helicity density follows the same form as in Eq. (28). As it is obvious from Eq. (30) the maximum of helicity enhancement $|h_{sca}/h_{inc}^o|$ occurs at points such $\theta = \pi/2$, on the surface of the NA, i.e., at $r = a$. Similar to what was shown in the previous subsection, we have depicted the total helicity enhancement $|h/h_{inc}^o|$ for the case of crystalline and amorphous silicon spheres in FIGs. 9 (a) and (b), respectively. In FIGs. 9 (c) and (d) we show maps of helicity density enhancement for a Si NA with radius $a = 80\,\text{nm}$, i.e., for the radius that generates maximum helicity enhancement at $\lambda = 680\,\text{nm}$ in the $x$-$y$ and $y$-$z$ planes, respectively. As observed from these figures, an enhancement of an order of magnitude is achieved around the sphere at points with $\theta = \pi/2$. Comparing FIG. 9 (c) and (d) with FIG. 8, we observe that there is a major difference between the results in this scenario and that of the ARPB illumination. Here, we have less enhancement while the area that the helicity is enhanced is more dispersed and is evenly distributed, i.e., is azimuthally uniform around the sphere over 360 degrees. However, in the ARPB illumination, the helicity is more localized and hence it is more enhanced. Indeed, depending on the application one may chose the most suitable illumination. For instance, if we are able to control the position of the chiral nanoparticle to locate it at a particular point, then, the ARPB illumination is more efficient. However, if we do not have a full control on the position of the chiral nanoparticle to be characterized, then, GB with circular polarization is more suitable.

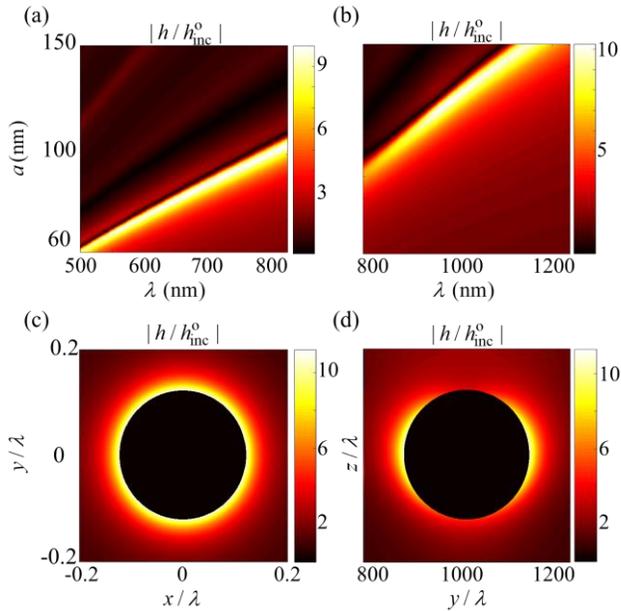

FIG. 9. Helicity density enhancement of the nearfield for various wavelengths and radii for (a) crystalline and (b) amorphous silicon NA under illumination of a circularly polarized GB with $w_0 = 3\lambda$. The map of helicity enhancement distribution is given in (c) for the $x$-$y$ plane, and in (d) for the $y$-$z$ plane, for the case of a crystalline silicon NA with radius $a = 80\,\mathrm{nm}$ at the wavelength $\lambda = 651\,\mathrm{nm}$.

## V. CONCLUSION

We have analyzed the interaction between small chiral nanoparticles and general electromagnetic fields and have determined the best field properties for chirality detection and characterization at nanoscale. We have demonstrated that any pair of electromagnetic fields with equal electric and magnetic energy densities is capable of detecting chirality if the pair constituents possess *unequal* helicity densities. Moreover, such pair of fields with equal electric and magnetic energy densities may be employed for chirality characterization only when the two fields possess opposite handedness and both satisfy the universal relation $|h| = u / \omega$ with $h$ and $u$ being the time-averaged helicity and energy densities of the field, respectively. In particular, we have investigated chiral fields of two kinds of optical beams: Gaussian beams with circular polarization, and combinations of beams with azimuthal symmetry forming an ARPB that needs to be properly engineered to satisfy the conditions in Eq. (11). Our analysis demonstrates that it is possible to enhance helicity density of chiral optical beams for the improvement of chirality detection by manipulating the relative phase, magnitude, and orientation of electric and magnetic fields. For example, we have shown an order of magnitude enhancement in helicity density for a circularly polarized GB and an ARPB. Moreover, we have illustrated that nearfields of an engineered high refractive index spherical NA is optimally chiral, i.e., satisfy conditions in Eq. (11), while localizing energy and helicity densities below diffraction limit and enhancing them for an order of magnitude.

Our study confirms the fundamental limit of energy density as a determinant factor for the upper bound of helicity density introduced in our concurrent work[1] and proves that such a bound is achievable when the fields are properly engineered. These findings enable the design of efficient field structures for chirality detection and characterization at nanoscale.

The authors acknowledge support by the W. M. Keck Foundation, USA.

References

[1] M. Hanifeh, M. Albooyeh, and F. Capolino, "Optimally Chiral Fields: Helicity Density and Interaction of Structured Light with Nanoscale Matter," *ArXiv1809 Phys.*

[2] J. P. Riehl, *Mirror-Image Asymmetry: An Introduction to the Origin and Consequences of Chirality.* John Wiley & Sons, 2011.

[3] R. J. Crossley, *Chirality and Biological Activity of Drugs.* CRC Press, 1995.

[4] D. B. Amabilino, *Chirality at the Nanoscale: Nanoparticles, Surfaces, Materials and More.* John Wiley & Sons, 2009.

[5] L. D. Barron, "From Cosmic Chirality to Protein Structure: Lord Kelvin's Legacy," *Chirality*, vol. 24, no. 11, pp. 879–893, Nov. 2012.

[6] G. H. Wagnière, *On Chirality and the Universal Asymmetry: Reflections on Image and Mirror Image.* John Wiley & Sons, 2008.

[7] B. A. Wallace and R. W. Janes, *Modern Techniques for Circular Dichroism and Synchrotron Radiation Circular Dichroism Spectroscopy.* IOS Press, 2009.

[8] E. U. Condon, "Theories of Optical Rotatory Power," *Rev. Mod. Phys.*, vol. 9, no. 4, pp. 432–457, Oct. 1937.

[9] I. Lindell, A. Sihvola, S. Tretyakov, and A. Viitanen, *Electromagnetic waves in chiral and bi-isotropic media.* Artech House, 1994.

[10] M. Oksanen and A. Hujanen, "How to Determine Chiral Material Parameters," in *1992 22nd European Microwave Conference*, 1992, vol. 1, pp. 195–199.

[11] S. Bassiri, C. H. Papas, and N. Engheta, "Electromagnetic wave propagation through a dielectric–chiral interface and through a chiral slab," *JOSA A*, vol. 5, no. 9, pp. 1450–1459, Sep. 1988.

[12] Y. Tang and A. E. Cohen, "Enhanced Enantioselectivity in Excitation of Chiral Molecules by Superchiral Light," *Science*, vol. 332, no. 6027, pp. 333–336, Apr. 2011.

[13] M. Schäferling, D. Dregely, M. Hentschel, and H. Giessen, "Tailoring Enhanced Optical Chirality: Design Principles for Chiral Plasmonic Nanostructures," *Phys. Rev. X*, vol. 2, no. 3, p. 031010, Aug. 2012.

[14] M. Schäferling, X. Yin, N. Engheta, and H. Giessen, "Helical Plasmonic Nanostructures as Prototypical Chiral Near-Field Sources," *ACS Photonics*, vol. 1, no. 6, pp. 530–537, Jun. 2014.

[15] N. Meinzer, E. Hendry, and W. L. Barnes, "Probing the chiral nature of electromagnetic fields surrounding plasmonic nanostructures," *Phys. Rev. B*, vol. 88, no. 4, p. 041407, Jul. 2013.

[16] A. O. Govorov and Z. Fan, "Theory of Chiral Plasmonic Nanostructures Comprising Metal Nanocrystals and Chiral Molecular Media," *ChemPhysChem*, vol. 13, no. 10, pp. 2551–2560, Jul. 2012.

[17] N. A. Abdulrahman *et al.*, "Induced Chirality through Electromagnetic Coupling between Chiral Molecular Layers and Plasmonic Nanostructures," *Nano Lett.*, vol. 12, no. 2, pp. 977–983, Feb. 2012.

[18] B. Auguié, J. L. Alonso-Gómez, A. Guerrero-Martínez, and L. M. Liz-Marzán, "Fingers Crossed: Optical Activity of a Chiral Dimer of Plasmonic Nanorods," *J. Phys. Chem. Lett.*, vol. 2, no. 8, pp. 846–851, Apr. 2011.

[19] B. Frank *et al.*, "Large-Area 3D Chiral Plasmonic Structures," *ACS Nano*, vol. 7, no. 7, pp. 6321–6329, Jul. 2013.

[20] A. O. Govorov, Z. Fan, P. Hernandez, J. M. Slocik, and R. R. Naik, "Theory of Circular Dichroism of Nanomaterials Comprising Chiral Molecules and Nanocrystals: Plasmon Enhancement, Dipole Interactions, and Dielectric Effects," *Nano Lett.*, vol. 10, no. 4, pp. 1374–1382, Apr. 2010.

[21] V. K. Valev, J. J. Baumberg, C. Sibilia, and T. Verbiest, "Chirality and Chiroptical Effects in Plasmonic Nanostructures: Fundamentals, Recent Progress, and Outlook," *Adv. Mater.*, vol. 25, no. 18, pp. 2517–2534, May 2013.

[22] D. Lin and J.-S. Huang, "Slant-gap plasmonic nanoantennas for optical chirality engineering and circular dichroism enhancement," *Opt. Express*, vol. 22, no. 7, pp. 7434–7445, Apr. 2014.

[23] G. Pellegrini, M. Finazzi, M. Celebrano, L. Duò, and P. Biagioni, "Surface-enhanced chiroptical spectroscopy with superchiral surface waves," *Chirality*, vol. 30, no. 7, pp. 883–889, Jul. 2018.

[24] W. Kuhn, "Optical Rotatory Power," *Annu. Rev. Phys. Chem.*, vol. 9, no. 1, pp. 417–438, 1958.

[25] J. D. Jackson, *Classical Electrodynamics, 3rd Edition*. 1998.

[26] A. Serdiukov, I. Semchenko, S. Tertyakov, and A. Sihvola, *Electromagnetics of bi-anisotropic materials - Theory and Application*, vol. 11. Gordon and Breach Science Publishers, 2001.

[27] M. Mansuripur, "Force, torque, linear momentum, and angular momentum in classical electr<Emphasis Type="Bold">odynamics</Emphasis>," *Appl. Phys. A*, vol. 123, no. 10, p. 653, Oct. 2017.

[28] D. Nowak *et al.*, "Nanoscale chemical imaging by photoinduced force microscopy," *Sci. Adv.*, vol. 2, no. 3, p. e1501571, Mar. 2016.

[29] M. Rajaei, M. A. Almajhadi, J. Zeng, and H. K. Wickramasinghe, "Near-Field Nanoprobing Using Si Tip-Au Nanoparticle Photoinduced Force Microscopy with 120:1 Signal-to-Noise Ratio, Sub-6-nm Resolution," *ArXiv180405993 Phys.*, Apr. 2018.

[30] F. Huang, V. A. Tamma, M. Rajaei, M. Almajhadi, and H. Kumar Wickramasinghe, "Measurement of laterally induced optical forces at the nanoscale," *Appl. Phys. Lett.*, vol. 110, no. 6, p. 063103, Feb. 2017.

[31] J. Zeng *et al.*, "In pursuit of photo-induced magnetic and chiral microscopy," *EPJ Appl. Metamaterials*, vol. 5, p. 7, 2018.

[32] J. Zeng *et al.*, "Sharply Focused Azimuthally Polarized Beams with Magnetic Dominance: Near-Field Characterization at Nanoscale by Photoinduced Force Microscopy," *ACS Photonics*, vol. 5, no. 2, pp. 390–397, Feb. 2018.

[33] M. Kamandi *et al.*, "Enantiospecific Detection of Chiral Nanosamples Using Photoinduced Force," *Phys. Rev. Appl.*, vol. 8, no. 6, p. 064010, Dec. 2017.

[34] M. Kamandi *et al.*, "Unscrambling Structured Chirality with Structured Light at Nanoscale Using Photoinduced Force," *ArXiv180506468 Phys.*, May 2018.

[35] Y. Martin, C. C. Williams, and H. K. Wickramasinghe, "Atomic force microscope–force mapping and profiling on a sub 100-Å scale," *J. Appl. Phys.*, vol. 61, no. 10, pp. 4723–4729, May 1987.

[36] E. Bayati, K. Oguichi, S. Watanabe, D. P. Winebrenner, and M. H. Arbab, "Terahertz time-domain polarimetry (THz-TDP) for measuring chirality," in *2017 42nd International Conference on Infrared, Millimeter, and Terahertz Waves (IRMMW-THz)*, 2017, pp. 1–1.

[37] Y. Tang and A. E. Cohen, "Optical Chirality and Its Interaction with Matter," *Phys. Rev. Lett.*, vol. 104, no. 16, p. 163901, Apr. 2010.

[38] R. P. Cameron, S. M. Barnett, and A. M. Yao, "Optical helicity, optical spin and related quantities in electromagnetic theory," *New J. Phys.*, vol. 14, no. 5, p. 053050, 2012.

[39] H. K. Moffatt, "The degree of knottedness of tangled vortex lines," *J. Fluid Mech.*, vol. 35, no. 1, pp. 117–129, Jan. 1969.

[40] L. Woltjer, "A Theorem on Force-Free Magnetic Fields," *Proc. Natl. Acad. Sci.*, vol. 44, no. 6, pp. 489–491, Jun. 1958.

[41] A. F. Ranada, "On the magnetic helicity," *Eur. J. Phys.*, vol. 13, no. 2, p. 70, 1992.

[42] J. L. Trueba and A. F. Rañada, "The electromagnetic helicity," *Eur. J. Phys.*, vol. 17, no. 3, p. 141, 1996.

[43] K. Y. Bliokh, A. Y. Bekshaev, and F. Nori, "Dual electromagnetism: helicity, spin, momentum and angular momentum," *New J. Phys.*, vol. 15, no. 3, p. 033026, 2013.

[44] F. Bernardeau, C. Grojean, and J. Dalibard, *Particle Physics and Cosmology: the Fabric of Spacetime: Lecture Notes of the Les Houches Summer School 2006*. Elsevier, 2007.

[45] M. Veysi, C. Guclu, and F. Capolino, "Vortex beams with strong longitudinally polarized magnetic field and their generation by using metasurfaces," *JOSA B*, vol. 32, no. 2, pp. 345–354, Feb. 2015.

[46] M. Veysi, C. Guclu, and F. Capolino, "Focused azimuthally polarized vector beam and spatial magnetic resolution below the diffraction limit," *JOSA B*, vol. 33, no. 11, pp. 2265–2277, Nov. 2016.

[47] L. D. Landau, J. S. Bell, M. J. Kearsley, L. P. Pitaevskii, E. M. Lifshitz, and J. B. Sykes, *Electrodynamics of Continuous Media*. Elsevier, 2013.

[48] C. R. Simovski and S. A. Tretyakov, "Model of isotropic resonant magnetism in the visible range based on core-shell clusters," *Phys. Rev. B*, vol. 79, no. 4, p. 045111, Jan. 2009.

[49] S. Campione, C. Guclu, R. Ragan, and F. Capolino, "Enhanced Magnetic and Electric Fields via Fano Resonances in Metasurfaces of Circular Clusters of Plasmonic Nanoparticles," *ACS Photonics*, vol. 1, no. 3, pp. 254–260, Mar. 2014.

[50] M. Darvishzadeh-Varcheie, C. Guclu, and F. Capolino, "Magnetic Nanoantennas Made of Plasmonic Nanoclusters for Photoinduced Magnetic Field Enhancement," *Phys. Rev. Appl.*, vol. 8, no. 2, p. 024033, Aug. 2017.

[51] A. Alù and N. Engheta, "The quest for magnetic plasmons at optical frequencies," *Opt. Express*, vol. 17, no. 7, pp. 5723–5730, Mar. 2009.

[52] S. Mühlig *et al.*, "Self-Assembled Plasmonic Core–Shell Clusters with an Isotropic Magnetic Dipole Response in the Visible Range," *ACS Nano*, vol. 5, no. 8, pp. 6586–6592, Aug. 2011.

[53] V. Ponsinet *et al.*, "Resonant isotropic optical magnetism of plasmonic nanoclusters in visible light," *Phys. Rev. B*, vol. 92, no. 22, p. 220414, Dec. 2015.

[54] C. Guclu, M. Veysi, and F. Capolino, "Photoinduced Magnetic Nanoprobe Excited by an Azimuthally Polarized Vector Beam," *ACS Photonics*, vol. 3, no. 11, pp. 2049–2058, Nov. 2016.

[55] M. Rahmani *et al.*, "Plasmonic Nanoclusters with Rotational Symmetry: Polarization-Invariant Far-Field Response vs Changing Near-Field Distribution," *ACS Nano*, vol. 7, no. 12, pp. 11138–11146, Dec. 2013.

[56] A. Vallecchi, M. Albani, and F. Capolino, "Collective electric and magnetic plasmonic resonances in spherical nanoclusters," *Opt. Express*, vol. 19, no. 3, pp. 2754–2772, Jan. 2011.

[57] D. Kajfez and P. Guillon, *Dielectric resonators*. 1986.

[58] O. Merchiers, F. Moreno, F. González, and J. M. Saiz, "Light scattering by an ensemble of interacting dipolar particles with both electric and magnetic polarizabilities," *Phys. Rev. A*, vol. 76, no. 4, p. 043834, Oct. 2007.

[59] I. Staude *et al.*, "Shaping Photoluminescence Spectra with Magnetoelectric Resonances in All-Dielectric Nanoparticles," *ACS Photonics*, vol. 2, no. 2, pp. 172–177, Feb. 2015.

[60] I. Sinev *et al.*, "Polarization control over electric and magnetic dipole resonances of dielectric nanoparticles on metallic films," *Laser Photonics Rev.*, vol. 10, no. 5, pp. 799–806, Sep. 2016.

[61] A. I. Kuznetsov, A. E. Miroshnichenko, M. L. Brongersma, Y. S. Kivshar, and B. Luk'yanchuk, "Optically resonant dielectric nanostructures," *Science*, vol. 354, no. 6314, p. aag2472, Nov. 2016.

[62] I. Staude *et al.*, "Tailoring Directional Scattering through Magnetic and Electric Resonances in Subwavelength Silicon Nanodisks," *ACS Nano*, vol. 7, no. 9, pp. 7824–7832, Sep. 2013.

[63] R. M. Bakker *et al.*, "Magnetic and Electric Hotspots with Silicon Nanodimers," *Nano Lett.*, vol. 15, no. 3, pp. 2137–2142, Mar. 2015.

[64] W. Liu, A. E. Miroshnichenko, D. N. Neshev, and Y. S. Kivshar, "Broadband Unidirectional Scattering by Magneto-Electric Core–Shell Nanoparticles," *ACS Nano*, vol. 6, no. 6, pp. 5489–5497, Jun. 2012.

[65] D. Permyakov *et al.*, "Probing magnetic and electric optical responses of silicon nanoparticles," *Appl. Phys. Lett.*, vol. 106, no. 17, p. 171110, Apr. 2015.

[66] Y. H. Fu, A. I. Kuznetsov, A. E. Miroshnichenko, Y. F. Yu, and B. Luk'yanchuk, "Directional visible light scattering by silicon nanoparticles," *Nat. Commun.*, vol. 4, p. 1527, Feb. 2013.

[67] B. S. Luk'yanchuk, N. V. Voshchinnikov, R. Paniagua-Domínguez, and A. I. Kuznetsov, "Optimum Forward Light Scattering by Spherical and

Spheroidal Dielectric Nanoparticles with High Refractive Index," *ACS Photonics*, vol. 2, no. 7, pp. 993–999, Jul. 2015.

[68] C.-S. Ho, A. Garcia-Etxarri, Y. Zhao, and J. Dionne, "Enhancing Enantioselective Absorption Using Dielectric Nanospheres," *ACS Photonics*, vol. 4, no. 2, pp. 197–203, Feb. 2017.

[69] A. García-Etxarri and J. A. Dionne, "Surface-enhanced circular dichroism spectroscopy mediated by nonchiral nanoantennas," *Phys. Rev. B*, vol. 87, no. 23, p. 235409, Jun. 2013.

[70] D. E. Aspnes and A. A. Studna, "Dielectric functions and optical parameters of Si, Ge, GaP, GaAs, GaSb, InP, InAs, and InSb from 1.5 to 6.0 eV," *Phys. Rev. B*, vol. 27, no. 2, pp. 985–1009, Jan. 1983.

[71] E. D. Palik, *Handbook of Optical Constants of Solids*. Academic Press, 2012.

[72] C. F. Bohren and D. R. Huffman, *Absorption and Scattering of Light by Small Particles*. John Wiley & Sons, 2008.

## APPENDIX A: LIGHT-MATTER INTERACTION

According to the Poynting theorem, the conservation of energy between a system of charged particles and electromagnetic fields implies [23]

$$\frac{1}{2}\int_V \nabla \cdot \Re\{\mathbf{E}\times\mathbf{H}^*\}d\boldsymbol{u} = -\frac{1}{2}\int_V \Re\{\mathbf{J}^*\cdot\mathbf{E}\}d\boldsymbol{u} + \\ \frac{\omega}{2}\int_V \Im\{\mathbf{E}^*\cdot\mathbf{D}+\mathbf{B}\cdot\mathbf{H}^*\}d\boldsymbol{u} \qquad (A1)$$

where we considered interaction of monochromatic electromagnetic fields and matters and averaged the quantities over one cycle of time. Here, $\mathbf{E}$, $\mathbf{B}$, $\mathbf{D}$, $\mathbf{H}$, and $\mathbf{J}$ are the "phasors" electric field, magnetic induction, electric displacement, magnetic field, and current density, respectively. Moreover, superscript "*" denotes conjugation and $V$ is the integration volume. We assume that the medium is non-conductive, i.e., $\mathbf{J}^*\cdot\mathbf{E} = 0$, and the constitutive relations $\mathbf{D} = \epsilon_0\mathbf{E}+\mathbf{P}$ and $\mathbf{B} = \mu_0\left(\mathbf{H}+\mathbf{M}\right)$ define the relations between fields and polarizations $\mathbf{P}$ and $\mathbf{M}$. Furthermore, the electric and magnetic dipoles of a nanoparticles of volume V are found via $\mathbf{p} = \int_V \mathbf{P}d\boldsymbol{u}$ and $\mathbf{m} = \int_V \mathbf{M}d\boldsymbol{u}$, i.e., by integrating the electric and magnetic volume polarizations. The electric and magnetic dipole moments are defined as $\mathbf{p} = \int_V \mathbf{r}\,\rho(\mathbf{r})d\boldsymbol{u}$ and $\mathbf{m} = \frac{1}{2}\int_V \mathbf{r}\times\mathbf{J}(\mathbf{r})d\boldsymbol{u}$, respectively. Note that here $\rho(\mathbf{r})$ is the electric charge density, not to be confused with the radial distance $\rho$ from $z$ axis introduced in section III. Therefore, Eq. (A1) reduces to

$$\int_V \nabla \cdot \frac{1}{2}\Re\{\mathbf{E}\times\mathbf{H}^*\}d\boldsymbol{u} = \frac{\omega}{2}\int_V \Im\{\mathbf{P}\cdot\mathbf{E}^*+\mu_0\mathbf{M}\cdot\mathbf{H}^*\}d\boldsymbol{u}. \qquad (A2)$$

The left and right sides of Eq. (A2) are the total power going out of the surface S surrounding the volume V and negative of the absorbed power in V, respectively. The extra negative sign for the absorbed power is considered because when calculating the absorbed power, we should integrate the power going into the volume V. The absorbed power $P_{\text{abs}}$ by an optically small particles reads

$$P_{\text{abs}} = \frac{\omega}{2}\Im\{\mathbf{p}^*\cdot\left(\mathbf{E}_{\text{loc}}+\mathbf{E}_{\text{sca}}\right)+\mu_0\mathbf{m}^*\cdot\left(\mathbf{H}_{\text{loc}}+\mathbf{H}_{\text{sca}}\right)\}, (A3)$$

where subscript "sca" and "loc" represent the scattered and local fields, respectively. The scattered nearfield is the nearfield generate by the nanoparticle, and the total field is the summation of scattered and incident (local) fields. By performing simple algebraic operations, the extinction power $P_{\text{ext}}$ and scattered power $P_{\text{sca}}$ are, respectively, obtained as

$$P_{\text{ext}} = \frac{\omega}{2}\Im\{\mathbf{p}\cdot\mathbf{E}_{\text{loc}}^*+\mu_0\mathbf{m}\cdot\mathbf{H}_{\text{loc}}^*\}, \qquad (A4)$$

and

$$P_{\text{sca}} = \frac{\omega k^3}{12\pi}\left(\epsilon_0^{-1}\,|\mathbf{p}|^2+\mu_0\,|\mathbf{m}|^2\right). \qquad (A5)$$

Note, in the calculation of scattered power in Eq. (A5), we have used the expressions (B1) and (B2) for the scattered electric and magnetic fields of an electric and a magnetic dipole, respectively.

## APPENDIX B: SCATTERING FIELDS OF A DIPOLE PARTICLE

For ease of reading we provide here the scattered electric and magnetic fields from the electric and magnetic dipole moments $\mathbf{p}_{\text{NA}}$ and $\mathbf{m}_{\text{NA}}$ of a NA located at the origin of the coordinate system [23]: the scattered nearfield is

$$\mathbf{E}_{\text{sca}} = \left\{(\hat{\mathbf{r}}\times\mathbf{p}_{\text{NA}})\times\hat{\mathbf{r}}-c_0^{-1}\left(1+\frac{i}{kr}\right)(\hat{\mathbf{r}}\times\mathbf{m}_{\text{NA}}) \right. \\ \left. +\left(\frac{1}{k^2r^2}-\frac{i}{kr}\right)\left[3\hat{\mathbf{r}}(\hat{\mathbf{r}}\cdot\mathbf{p}_{\text{NA}})-\mathbf{p}_{\text{NA}}\right]\right\}\frac{k^2e^{ikr}}{4\pi r\epsilon_0}, \qquad (B1)$$

and the magnetic field is

$$\mathbf{H}_{\mathrm{sca}} = \left\{ \left(\hat{\mathbf{r}} \times \mathbf{m}_{\mathrm{NA}}\right) \times \hat{\mathbf{r}} + c_0 \left(1 + \frac{i}{kr}\right) \left(\hat{\mathbf{r}} \times \mathbf{p}_{\mathrm{NA}}\right) + \right.$$

$$\left. \left(\frac{1}{r^2 k^2} - \frac{i}{rk}\right) \left[3\hat{\mathbf{r}}\left(\hat{\mathbf{r}} \cdot \mathbf{m}_{\mathrm{NA}}\right) - \mathbf{m}_{\mathrm{NA}}\right] \right\} \frac{k^2 e^{ikr}}{4\pi r}, \tag{B2}$$

where $r$ is the radial distance from the dipoles to the observation point, and $\hat{\mathbf{r}}$ is the unit position vector showing the direction from origin to the observation point. Therefore, the dominant terms in Eq. (B1) and (B2), when very close to the antenna (the nearfield) are obtained by retaining only the field terms with $1/r^3$, namely

$$\mathbf{E}_{\mathrm{sca}} \approx \frac{e^{ikr}}{4\pi\epsilon_0 r^3} \left[3\hat{\mathbf{r}}\left(\hat{\mathbf{r}} \cdot \mathbf{p}_{\mathrm{NA}}\right) - \mathbf{p}_{\mathrm{NA}}\right],$$

$$\mathbf{H}_{\mathrm{sca}} \approx \frac{e^{ikr}}{4\pi r^3} \left[3\hat{\mathbf{r}}\left(\hat{\mathbf{r}} \cdot \mathbf{m}_{\mathrm{NA}}\right) - \mathbf{m}_{\mathrm{NA}}\right]. \tag{B3}$$

As a result, the helicity density $h = \Im\{\mathbf{E} \cdot \mathbf{H}^*\}/(2\omega c_0)$ corresponding to the scattered nearfields of the NA takes the form of Eq. (23).

APPENDIX C

The $z$ component of the magnetic field of a circularly polarized Gaussian beam with electric field defined in Eq. (17) is obtained by applying Maxwell equation $\mathbf{H} = (i\omega\mu_0)^{-1}\nabla\times\mathbf{E}$ leading to

$$H_z = \frac{-2V_G}{\omega\mu_0} \left(\frac{ik}{2R} - \frac{1}{w^2}\right) \left(\frac{\rho}{w}\right) e^{i\Phi}, \tag{C1}$$

with $\quad \Phi = i(\rho/w)^2 + k\rho^2/(2R) - \tan^{-1}(z/z_R) + kz - \varphi$. This component is negligible when $\rho \ll w$ or in other words when $\rho \ll w_0$. Therefore, when $\rho \ll w_0$ the field of a chiral Gaussian beam locally resembles that of a plane wave with circular polarization.